\documentclass[twocolumn,aps,prb,10pt,superscriptaddress]{revtex4-2}

\usepackage{amsmath}
\usepackage{amssymb}
\usepackage{mathtools}
\usepackage{color}

\newcommand{\inb}[3]{{\left#1 #2 \right#3}}
\newcommand{\reff}[1]{{Fig.\ref{fig:#1}}}
\newcommand{\bs}[1]{\boldsymbol{#1}}
\newcommand{\bigtriangleleft}{\raisebox{-1.5pt}{$\scalebox{1.6}{$\triangleleft$}$}}
\newcommand{\bigtriangleright}{\raisebox{-1.5pt}{$\scalebox{1.6}{$\triangleright$}$}}
\newcommand{\squad}{\ }

\begin{document}

\title{Topological edge states in two-dimensional $\mathbb{Z}_4$ Potts paramagnet protected by the $\mathbb{Z}_4^{\times 3}$ symmetry
}

\author{Hrant Topchyan}
\affiliation{A.Alikhanyan National Science Laboratory (Yerevan Physics Institute), Yerevan 0036, Armenia}

\author{Tigran Hakobyan}
\affiliation{A.Alikhanyan National Science Laboratory (Yerevan Physics Institute), Yerevan 0036, Armenia}
\affiliation{Yerevan State University, Yerevan 0025, Armenia}

\author{Mkhitar Mirumyan}
\affiliation{A.Alikhanyan National Science Laboratory (Yerevan Physics Institute), Yerevan 0036, Armenia}

\author{Tigran A. Sedrakyan}
\affiliation{A.Alikhanyan National Science Laboratory (Yerevan Physics Institute), Yerevan 0036, Armenia}
\affiliation{Department of Physics, University of Massachusetts, Amherst, Massachusetts 01003, USA}

\author{Ara Sedrakyan}
\affiliation{A.Alikhanyan National Science Laboratory (Yerevan Physics Institute), Yerevan 0036, Armenia}

\begin{abstract}
We construct a two-dimensional bosonic symmetry-protected topological (SPT) paramagnet protected by an on-site $G=\mathbb{Z}_4^{\times 3}$ symmetry, starting from a three-component $\mathbb{Z}_4$ Potts paramagnet on a triangular lattice. Within the group-cohomology framework, $H^{3}(G,U(1))\cong \mathbb{Z}_4^{\times 7}$, we focus on a ``colorless" cocycle representative obtained by antisymmetrizing the basic $\mathbb{Z}_4$ three-cocycle, and generate the corresponding SPT Hamiltonian via a cocycle-induced local unitary transformation followed by symmetry averaging. For open geometry, we derive the boundary theory explicitly: one color sector decouples, while the nontrivial edge reduces to an interacting $\mathbb{Z}_4$ chain with next-nearest-neighbor constraints that admits a compact dressed-Potts form. Using DMRG we show that the boundary model is gapless, with the lowest gap scaling as $1/L$ and an entanglement-entropy scaling consistent with a conformal field theory of central charge $c=2.191(4)\simeq 11/5$. The rational value $c=11/5$ matches the coset $SU(3)_3/SU(2)_3$, making it a candidate for the continuum description of the $\mathbb{Z}_4^{\times 3}$ edge; we outline spectral and symmetry-resolved diagnostics needed to test this identification at the level of conformal towers beyond the central charge.
\end{abstract}

\maketitle

\section{Introduction}

The structure of quantum states of matter at zero temperature has become a central theme in condensed matter physics, motivated in part by rapid advances in quantum science and technology \cite{wen2019,Sachdev-2023}.
A key lesson is that the classical Landau framework---based on spontaneous symmetry breaking and local order parameters---does not exhaust the possible organizing principles of quantum phases \cite{Landau-1937,Landau-Lifshitz}.
In particular, two gapped ground states with the same microscopic symmetry
can represent distinct phases even in the absence of local order parameters.
For SPT phases, any path that connects such a state to a trivial product state
must either break the protecting symmetry or close the bulk gap.

Symmetry-protected topological (SPT) phases are among the simplest examples of such non-Landau quantum matter.
They are short-range entangled, gapped phases whose distinction is stable only in the presence of a protecting on-site global symmetry \cite{Gu-Wen-2009,Senthil-2015,Wen-Colloquium-2017,Berg-2009,Berg-2010,Wen-2011d,Kitaev-2011,Cirac-2011}.
Once the protecting symmetry is explicitly broken, different SPT phases can be adiabatically connected to a trivial product state without closing the bulk gap.
Topological insulators \cite{hasan-kane,qi-zhang} and the Haldane spin-$1$ chain \cite{haldane} provide canonical examples.

A defining property of an SPT phase is the constrained dynamics at its boundary.
While the bulk can be fully gapped and nondegenerate, a symmetry-preserving boundary generically cannot be both gapped and nondegenerate: it must either remain gapless, spontaneously break the symmetry, or develop intrinsic topological order.
This obstruction is captured by an anomalous realization of the symmetry at the boundary, i.e.\ the boundary symmetry cannot be implemented as a strictly on-site unitary action.
Equivalently, the bulk admits a symmetry-protected topological response whose anomaly inflow cancels the boundary anomaly \cite{Wen-2012,tHooft-1980}.
The anomaly thus provides a sharp bulk invariant and underlies the bulk--boundary correspondence.

For bosonic systems with on-site symmetry $G$ in $d$ spatial dimensions, the group-cohomology construction classifies a large family of SPT phases by $H^{d+1}(G,U(1))$ \cite{Wen-2012,Wen-2013}.
In $d=1$, the classification reduces to projective representations that appear at the ends of an open chain \cite{Wen-2011b,Wen-2011c}.
In $d=2$, the anomaly can be viewed as a failure of strict associativity of the symmetry action on the boundary; it is naturally exposed in the fusion of symmetry defect lines and is encoded by a three-cocycle in $H^{3}(G,U(1))$, often written in terms of $F$-symbols \cite{kawagoe-levin, Maeda2025}.
Related viewpoints connect certain SPT models to intrinsically topologically ordered phases by duality transformations \cite{levin-gu}.

Recently, both the conceptual and practical toolkits for boundary anomalies have expanded substantially.
On the conceptual side, generalized (including non-invertible and subsystem) symmetries and their symmetry topological field theory and topological-holography descriptions provide a unifying language for anomalies, gapped boundaries, and (intrinsically) gapless symmetry-protected phases \cite{thorngren-wang-fusion-category-2024,seiberg-seifnashri-shao-noninv-2024,bhardwaj-gapped-noninv-2025,antinucci-symtft-gapless-spt-2025,jia-jia-subsystem-symtft-2025,seiberg-shao-zhang-lsm-cpt-2025,Seiberg2025, KWTS1, KWTS2}.
On the practical side, systematic analyses of symmetry-preserving boundary conditions in CFT and refined numerical diagnostics of boundary anomalies continue to sharpen the bulk--boundary correspondence \cite{li-hsieh-yao-oshikawa-prb-2024,oshikawa-scaling,ding-sspt-anomaly-2025,loo-wang-fspt-interfaces-2025,xu-jian-average-exact-2025,guo-yang-double-aspt-2025}.

Concrete realizations of bosonic SPT phases via exactly solvable Hamiltonians with commuting local terms include decorated-domain-wall constructions \cite{Vishwanath-2013}, cluster models \cite{Wen-2011b,Yoshida-2016-1}, the Levin-Gu model \cite{levin-gu} for $\mathbb{Z}_2$ symmetry, and its generalizations to many-body states\cite{SGK,CSSC, WWS} and $\mathbb{Z}_3^{\times 3}$ and $\mathbb{Z}_3$ symmetries \cite{spt_z33,spt_z2f,spt_z3}.
Related decorated-defect ideas have also been adapted to construct gapless SPT phases with protected boundary signatures \cite{li-decorated-gapless-spt-2024}.

In the seminal work \cite{levin-gu}, Levin and Gu showed that the nontrivial $(2+1)$-dimensional bosonic $\mathbb{Z}_2$ SPT admits no symmetry-preserving trivially gapped edge.
Instead, the boundary degrees of freedom can be captured by a critical one-dimensional XX spin chain, i.e.\ the gapless point of the XY model.
For the $\mathbb{Z}_3$ and $\mathbb{Z}_3^{\times 3}$ cases, Refs.~\cite{spt_z33,spt_z2f,spt_z3} constructed exactly solvable bulk Hamiltonians and derived effective boundary Hamiltonians.
In the continuum, the low-energy edge theory was argued to be described by the coset CFT $SU_k(3)/SU_k(2)$ at level $k$.
Numerical analysis suggested that the $\mathbb{Z}_3$ edge corresponds to $k=1$, while the $\mathbb{Z}_3^{\times 3}$ edge corresponds to $k=2$.

In this work, we extend the construction of Refs.~\cite{spt_z33,spt_z2f,spt_z3} to the paramagnetic $\mathbb{Z}_4^{\times 3}$-symmetric Potts model on a triangular lattice and derive the associated effective boundary Hamiltonian.
We then study its spectrum and entanglement properties to identify the emergent boundary CFT data.

\section{The $\mathbb{Z}_4^{\times 3}$ SPT model}

In this section, we study  the  paramagnetic Potts model on a triangular lattice
where the on-site states are given by the elements of the threefold product of the cyclic group of order four,
\begin{equation}
\label{G}
G=\mathbb{Z}_4^{\times3}=\mathbb{Z}^A_4\times \mathbb{Z}^B_4\times \mathbb{Z}^C_4\squad.
\end{equation}
Here, $\alpha\in\{A,B,C\}$ is the ``color" index associated with the first, second, and third
cyclic groups within the product, respectively.

Then the elements of group $G$ can be given
as 3-vectors with integer components (additive representation)
and the group operation is the component-wise addition by modulo 4:
\begin{equation}
\begin{aligned}
&\bs{n}=(n^A, n^B, n^C)\in G
\squad\text{with}\squad
n^\alpha\in\{0,1,2,3\} \squad,\\
&\bs{n}_1+\bs{n}_2=\{n^\alpha_1+n^\alpha_2\mod 4\} \squad.
\end{aligned}
\label{n}
\end{equation}

The lattice can be constructed from a three-colored triangular lattice with four-state
sites by merging $\nabla$-type triangular faces into single composite sites in a specific way,
as illustrated in \reff{tritri}.
Then the on-site state of the constructed lattice  is labeled by the integral vectors \eqref{n}.
For a given color $\alpha$, one can define operators
\begin{equation}
\begin{split}
&X\inb|{n}> = \inb|{n - 1 \text{~mod~} 4}>\squad,\quad
Z\inb|{n}> = i^{n} \inb|{n}>\squad,\\
&X Z = i Z X\squad,\quad
X^4 = Z^4 = 1\squad.
\end{split}
\label{XZ}
\end{equation}
The operators at different points commute.

The noninteracting (paramagnetic) Potts model,
which we use as the base for SPT phase construction,
is described by the Hamiltonian
\begin{equation}
\label{H_0}
	H_0 = -\sum_p \sum_\alpha \Big(X^\alpha_p + (X^\alpha_p)^2 + (X^\alpha_p)^3\Big)\squad,
\end{equation}
where the first sum is performed over the sites $p$ of the constructed lattice
and the second is over colors $A,B,C$.
\begin{figure}
	\centering
	\includegraphics[width = .8\linewidth]{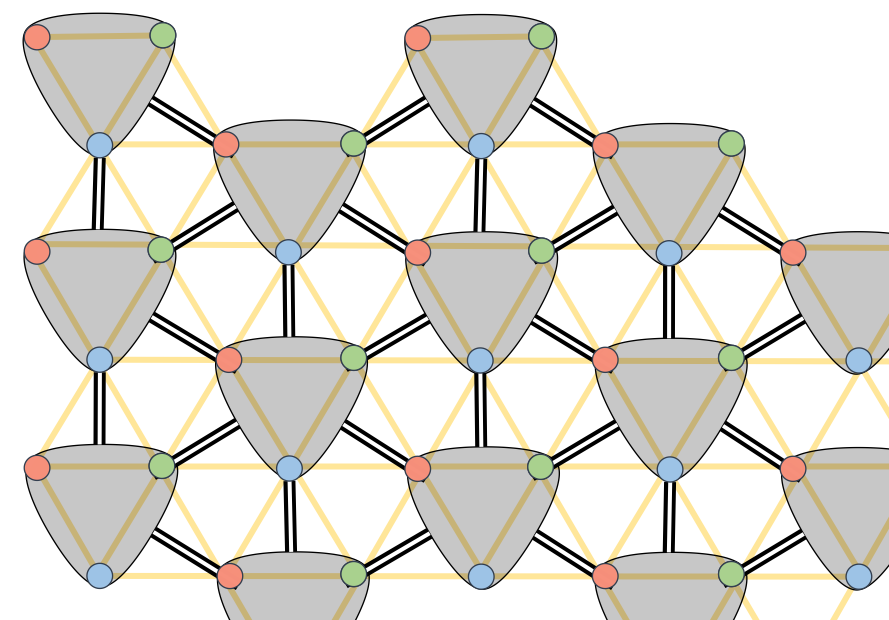}
	\caption{
        The initial $\mathbb{Z}_4$ triangular lattice
        with colored (red, green, blue) nodes and yellow links and
        the colorless composite $\mathbb{Z}_4^{\times 3}$ triangular lattice with
        gray-shaded triangles as the nodes and
		the black double line links.}
	\label{fig:tritri}
\end{figure}
The system \eqref{H_0} has a global symmetry $G$
in representation $S_{\bs{g}}$ ($\bs{g}\in G$),
\begin{equation}
\label{Sg}
[H_0, S_{\bs{g}}]=0\squad,\quad
 S_{\bs{g}} = \prod_{p} \prod_\alpha (X_p^\alpha)^{g^\alpha} \squad.
\end{equation}

\subsection{General structure of the SPT phase space}

According to general theory, the nontrivial SPT phases of two-dimensional models
are classified  by the nontrivial elements of
the third $U(1)$ cohomology of the symmetry group.
In our case, the latter is the sevenfold product
of the underlying cyclic group \cite{cohoms}:
\begin{equation}
\label{cohom}
H^3(\mathbb{Z}_4^{\times 3}, U(1))= \mathbb{Z}_4^{\times 7}\squad.
\end{equation}

The $k$-th cohomology group is defined as the factor of
$k$-``cocycles" (or closed forms) to $k$-``coboundaries" (or trivial forms).
$k$-cocycles and $k$-coboundaries 
of a group $G$ are the $G$-invariant functions (cochains, forms)
$\nu_k:G^{\times (k+1)} \rightarrow U(1)$,
$\nu_k(g_0, \dots, g_k)=\nu_k(g+g_0, \dots, g+g_k)$
that obey conditions $\delta\nu_k = 1$ and $\nu_k = \delta\nu_{k-1}$, respectively.
Here, $\delta : \nu_{k-1}\rightarrow \nu_k$ is the ``coboundary" operator,
which is defined in a way to exhibit a structure analogous to that
of the standard exterior derivative of differential forms.
Similarly, it is also nilpotent: $\delta^2 \nu_k =1$.
The relevant expression of $\delta$ is
\begin{equation}
\label{delta}
\delta\nu_3(\bs{n}_0,\dots,\bs{n}_4)=
\prod_{i=0}^4 \nu_3^{(-1)^i}(\bs{n}_0, \dots, \check{\bs{n}}_i, \dots, \bs{n}_4)
\end{equation}
where notation $\check{\bs{n}}_i$ indicates that the $i$-th argument is omitted.
Using the $G$-invariance, one can eliminate the first argument of the form,
yielding an alternative representation
\begin{equation}
\omega_3(\bs{n}_1, \bs{n}_2, \bs{n}_3)
=\nu_3(0,\bs{n}_1,\bs{n}_1+\bs{n}_2,\bs{n}_1+\bs{n}_2+\bs{n}_3)\squad,
\label{w-nu}
\end{equation}
\begin{equation}
\begin{split}
&\delta\omega_3(\bs{n}_1,\dots,\bs{n}_4)\\
=&\frac{\omega_3(\bs{n}_2,\bs{n}_3,\bs{n}_4)\omega_3(\bs{n}_1,\bs{n}_2+\bs{n}_3,\bs{n}_4) \omega_3(\bs{n}_1,\bs{n}_2,\bs{n}_3)}
{\omega_3(\bs{n}_1+\bs{n}_2,\bs{n}_3,\bs{n}_4)\omega_3(\bs{n}_1,\bs{n}_2,\bs{n}_3+\bs{n}_4)}
\squad.\hspace{-8pt}
\label{delta-w}
\end{split}
\end{equation}
The relations \eqref{delta}, \eqref{w-nu}, and \eqref{delta-w}
extend naturally to arbitrary dimensions.
Consequently, the closedness condition $\delta \omega_3= 1$
can be verified directly.
In contrast, establishing the exactness condition is more subtle.
In \cite{spt_z33}, it was established that if the 3-cocycle
is trivial ($\omega_3 = \delta \omega_2$ for some 2-cochain $\omega_2$)
then it satisfies the equation
\begin{align}
\label{R}
R[\omega_3]:=\prod_{\sigma\in\mathbb{S}_3}
\omega_3^{\epsilon(\sigma)}(\bs{n}_{\sigma_1}, \bs{n}_{\sigma_2}, \bs{n}_{\sigma_3})
\xrightarrow[\omega_3=\delta \omega_2]{} 1 \squad.
\end{align}
Here, the product runs over all permutations
$\sigma=\{\sigma_1, \sigma_2, \sigma_3\}\in \mathbb{S}_3$ of the set $\{1,2,3\}$
and $\epsilon(\sigma)=\pm 1$ is the parity of the permutation $\sigma$.

\subsection{Construction of nontrivial SPT phases}

The nontrivial SPT phases  are described by Hamiltonians
obtained from the original model  \eqref{H_0} by a local unitary transformation
generated by nontrivial 3-cocycles  $\nu_3\in H^3(G,U(1))$ with
subsequent symmetrization over the symmetry group $G$
\cite{Yoshida-2016-1,Yoshida-2017,spt_z33, spt_z3, spt_z2f}:
\begin{equation}
\label{H}
\begin{split}
	H &=\frac1{|G|} \sum_{\bs{g}\in G} S_{\bs{g}} U H_0 U^\dagger S_{\bs{g}}^\dagger
    \squad,\\
   U &= \prod_{\inb<{pqr}>} \nu_3^{\epsilon_{\inb<{pqr}>}}(0,\bs{n}_p,\bs{n}_q,\bs{n}_r)\squad,
\end{split}
\end{equation}
where the product is taken over all triangular faces of the composite lattice. The
sign factor $\epsilon_\inb<{pqr}>=\pm1$ differentiates the triangle orientations:
  $\epsilon=1$ for the triangles pointing left (\bigtriangleleft)
  and $\epsilon=-1$ for those pointing right (\bigtriangleright).
 $|G|=64$ is the order of the symmetry group \eqref{G}.

Without the symmetrization procedure in \eqref{H},
$H_0$ and $H$ would be unitarily equivalent.
This is also the case for a closed system, even with symmetrization, as
$[S_{\bs{g}}, U]=0$ ($g\in G$) resulting from of $\delta \nu_3 = 0$ \cite{spt_z2f}.
In the general case of an open system $[S_{\bs{g}}, U] \neq 0$,
but rather $S_{\bs{g}} U = V_{\partial, \bs{g}} U S_{\bs{g}}$
where $V_{\partial, \bs{g}}$ are some terms on the system's boundary links
(which makes it trivial on a closed system as $\partial=\varnothing$). 
The averaging procedure over $G$ in \eqref{H} is meant to restore the broken symmetry.
If one separates the bulk ($B$) and boundary ($\partial$) sectors of the Hamiltonians,
$H_0=H_{0,B}+H_{0,\partial}$, $H=H_B+H_\partial$,
since $[S_{\bs{g}}, H_0]=0$ and by definition $[V_{\partial, \bs{g}}, H_{0,B}]=0$,
then $H_B$ will be $U H_{0,B} U^\dagger$
while $H_\partial$ will contain additional terms $V_{\partial, \bs{g}}$ compared to $U H_{0,\partial} U^\dagger$.
It is also apparent that $[H_B,H_\partial]=0$ \cite{Yoshida-2017, spt_z2f}.

Let us now return to the cohomology group \eqref{cohom} of the considered model,
which consists of seven independent $\mathbb{Z}_4$ generators \cite{spt_z33}.
The corresponding forms are well known and their color structure has been
identified (see \cite{mark1995} as an example).
Three of these generators act only on one specific color, $A$, $B$ or $C$,
and are the same as the cohomology $H^3(\mathbb{Z}_4,U(1))=\mathbb{Z}_4$
generator of a single group $\mathbb{Z}_4$.
The resulting phases are therefore identical to the SPT phases
protected by a single $\mathbb{Z}_4$ symmetry (type I phases).
In contrast, three other generators act on color pairs, $\{A,B\}$, $\{B,C\}$ and $\{A,C\}$.
The associated type-II phases encode pairwise mixed anomalies between the corresponding $\mathbb{Z}_4$ charges and are not the focus of the present work.
However, none of the six aforementioned generators utilizes the whole Hilbert space,
and the corresponding phases can be realized in a system with a smaller symmetry.
Equivalently, the three single-color generators describe pure one-color boundary anomalies,
whereas the three pair generators describe mixed anomalies involving two colors and leave the
third color as a spectator.  The remaining generator is the genuine three-color, or type-III,
component. It is the only one whose boundary obstruction necessarily involves all three
$\mathbb{Z}_4$ charges simultaneously.

The single ``colorless" $\mathbb{Z}_4$ component is generated by the cocycle
\begin{equation}
\label{ws}
	\omega_3^o(\bs{n}_1, \bs{n}_2, \bs{n}_3) = i^{n_1^A n_2^B n_3^C}\squad.
\end{equation}
Its closedness can be verified directly and
nontriviality is confirmed using \eqref{R}
by verifying that, for instance,
\begin{equation}
\label{Rws}
	R[\omega_3^o]((1, 0, 0), (0, 1, 0), (0, 0, 1)) = i\ne 1\squad.
\end{equation}
$\omega_3^o$ is a representative of the
cocycle class that corresponds to some element $q$
of the factor (cohomology) group,
and specifically the ``colorless" component (direct factor) $\mathbb{Z}_4$.
It is easy to show that $q$ is the generator of that group:
obviously, $\delta (\omega_3 \omega_3') = \delta \omega_3  \delta \omega_3'$ \eqref{delta} and
$R[\omega_3 \omega_3'] = R[\omega_3] R[\omega_3']$ \eqref{R}.
$(\omega_3^o)^2$ and $(\omega_3^o)^3$,
which correspond to the group elements $q^2$ and $q^3$, respectively,
are thus nontrivial cocycles as well.
Since all $q$, $q^2$ and $q^3$ are nontrivial, $q$ is the generator
of the cohomology group component $\mathbb{Z}_4$,
and the different degrees of $\omega_3^o$ cover the whole space of
substantially different cocycles.

In this article, we consider a derived cohomology element obtained
by antisymmetrizing \eqref{ws} over its last two arguments: 
\begin{equation}
\label{wa}
	\omega_3^a(\bs{n}_1, \bs{n}_2, \bs{n}_3) = i^{n_1^A \inb({n_2^B n_3^C - n_2^C n_3^B})}\squad.
\end{equation}
The closedness condition $\delta \omega_3^a=1$ is easily verified,
and the sufficient condition for the cocycle to be nontrivial $R[\omega_3^a]\neq 1$
can be easily checked by evaluation with the same arguments as in \eqref{Rws}.
Moreover, one can check that the counterpart $\omega_3^c=i^{-n_1^A n_2^C n_3^B}$
that was added to $\omega_3^o$ belongs to the same cohomology as $\omega_3^o$ itself
($\omega_3^c \equiv \omega_3^o$).
Indeed, $\omega_3^o/\omega_3^c = \delta \omega_2$
with $\omega_2(\bs{n}_1 \bs{n}_2)=i^{n_1^A n_2^B n_2^C}$.
Thus $\omega_3^a=\omega_3^o \omega_3^c\equiv(\omega_3^o)^2$, so $\omega_3^a$
represents the order-two element of the ``colorless" cohomology component $\mathbb{Z}_4$.

The corresponding invariant cocycle  can be reconstructed using the relation \eqref{w-nu}:
%
\begin{equation}
\begin{split}
	\nu_3(0, \bs{s}, \bs{x}, \bs{y}) =
	\omega_3^a(\bs{s}, \bs{x}-\bs{s}, \bs{y}-\bs{x})
    &= i^{\varphi_3(\bs{s},\bs{x},\bs{y})} \squad,\\
	\varphi_3 (\bs{s}, \bs{x},\bs{y}) = s^A \big[x^B y^C - x^C y^B
	-&s^B \inb({y^C - x^C}) \\ + &s^C \inb({y^B - x^B})\big] \squad.
\end{split}
\label{nu}
\end{equation}

\subsection{The boundary model}

Let us apply the general expression of the SPT Hamiltonians \eqref{H}
to this particular cocycle. 
Note that the exponent $\psi_3$ in \eqref{nu} is antisymmetric
with regard to the permutation of its last two arguments, so
\begin{equation}
\label{nu-asym}
\nu_3(0, \bs{s}, \bs{x}, \bs{y})=\nu_3(0, \bs{s}, \bs{y}, \bs{x})^{-1}\squad.
\end{equation}
This condition is required in order for the edge Hamiltonian to be
translation-invariant and independent of the specific shape of the edge \cite{spt_z2f}.
\begin{align}
    H_\partial=-\frac{1}{64} \sum_s \sum_{p \in \partial, \alpha}
	&V_{\bs{s},p} \Big(X_p^\alpha + (X_p^\alpha)^2 + (X_p^\alpha)^3\Big) V_{\bs{s},p}^{-1}
    \nonumber \\ \text{with} \quad &V_{\bs{s},p} =
	\frac{\nu_3(0,-\bs{s},\bs{n}_p,\bs{n}_{p-1})}{\nu_3(0,-\bs{s},\bs{n}_p,\bs{n}_{p+1})}\squad.
    \label{V2H}
\end{align}
One may observe that any terms of the form $f_{\bs{s}}(\bs{x})/f_{\bs{s}}(\bs{y})$
appearing in $\nu_3(0,-\bs{s},\bs{x},\bs{y})$ cancel out in \eqref{V2H},
so such terms can be omitted from the outset.
More precisely, if there is
$\nu'(0,-\bs{s},\bs{x},\bs{y})=\nu(0,-\bs{s},\bs{x},\bs{y})
\cdot f_{\bs{s}}(\bs{x}) / f_{\bs{s}}(\bs{y})$,
then the corresponding $V_{\bs{s},p}'=V_{\bs{s},p} \cdot
f_{\bs{s}}(\bs{n}_{p+1}) / f_{\bs{s}}(\bs{n}_{p-1})$.
The additional terms are commutative with $X_p^\alpha$,
so they vanish in the transformation $V_{\bs{s},p}(\cdots)V_{\bs{s},p}^{-1}$,
thus allowing for such $\nu \rightarrow \nu'$ substitutions
for the purpose of deriving $H_\partial$.

The expressions \eqref{nu} are then simplified to
\begin{equation}
\label{nu'}
\nu_3'(0,-\bs{s},\bs{x},\bs{y}) = i^{-s^A ({x^B y^C - x^C y^B})}\squad.
\end{equation}
Using the definition \eqref{XZ},
it is straightforward to verify the relations
\begin{equation}
\nonumber
Z_p=i^{n_p}\squad,\squad
X_p^\alpha i^{n_p^\beta} = i^{n_p^\beta + \delta_\alpha^\beta} X_p^\alpha\squad,\squad
\sum_{s=0}^3 i^{ks} = 4\delta^{(4)}_k\squad,
\end{equation}
where $\delta^{(N)}$ denotes the Kronecker delta modulo $N$,
$\delta^{(N)}_k=\delta_{k~\text{mod}~N}$.
Then the similarity transformations in the expression for the edge Hamiltonian \eqref{V2H}
reduce the local terms as follows:
\begin{align}
(X_p^A)^k &\rightarrow (X_p^A)^k\squad,\label{local}\\
(X_p^B)^k &\rightarrow \frac{1}{4} (X_p^B)^k \sum_{s=0}^3 \inb({Z_{p-1}^C Z_{p+1}^C})^{ks}
\nonumber=(X_p^B)^k \delta_{k\Delta n_p^C}^{(4)}\squad,\nonumber\\
(X_p^C)^k &\rightarrow \frac{1}{4} (X_p^C)^k \sum_{s=0}^3 \inb({Z_{p-1}^B Z_{p+1}^B})^{ks}
\nonumber=(X_p^C)^k \delta_{k\Delta n_p^B}^{(4)}\squad.\nonumber
\end{align}
where we introduced the notations $\Delta n_p^\alpha = n_{p+1}^\alpha - n_{p-1}^\alpha$.

\begin{figure}
	\centering
	\includegraphics[width = .75\linewidth]{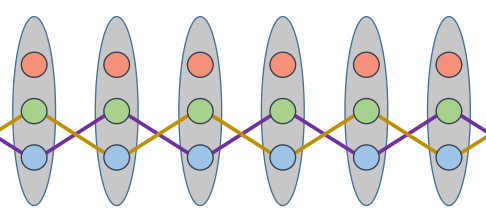}
	\caption{The schematic depiction of interactions.
		The gray-shaded blobs are the lattice edge points,
		with their internal components of colors $A$ (red),
		$B$ (green) and $C$ (blue).
		The interacting components are connected by brown or purple lines.}
	\label{fig:edge_struct}
\end{figure}


The boundary Hamiltonian $H_\partial$ contains
next-nearest-neighbor interactions,
schematically shown in \reff{edge_struct}.
One can see a trivial separation of the color $A$
and the splitting of the remaining terms into two distinct $\mathbb{Z}_4$ chains,
each consisting of the colors $B$ and $C$  in alternating order.
This occurs when the edge has even length; otherwise, the seemingly separate chains
combine into a single chain that is twice as long.

Using the relations $\delta^{(4)}_{2n}=\delta^{(2)}_{n}$ and
$\delta^{(4)}_{3n}=\delta^{(4)}_{n}$ for Kronecker deltas
to simplify the local terms \eqref{local},
one can acquire the Hamiltonian of a single $\mathbb{Z}_4$ chain in the form
\begin{equation}
	\widetilde{H}_\partial =
	-\sum_p \left[(X_p + X_p^\dagger) \delta_{\Delta n_p}^{(4)}
		+ X_p^2 \delta_{\Delta n_p}^{(2)}\right] \squad,
	\label{eq:ham_init}
\end{equation}
with $\Delta n_p = n_{p+1} - n_{p-1}$.
The color labels disappear at this stage,
since we now deal with a single
$\mathbb{Z}_4$ chain rather than a $\mathbb{Z}_4^{\times 3}$ system.
The tilde in $\widetilde{H}_\partial$ indicates that
it is only the nontrivial part of $H_\partial$.

The Hamiltonian \eqref{eq:ham_init} is the $N=4$ member of an explicit
$\mathbb{Z}_N^{\times 3}$ family \cite{spt_zN3}.
For the analogous $N$-state Potts paramagnet,
the antisymmetric type-III cocycle corresponding to \eqref{ws} is
\begin{equation}
\omega_{3,N}^{a}(\bs{n}_1,\bs{n}_2,\bs{n}_3)
=\zeta_N^{\,n_1^A(n_2^B n_3^C-n_2^C n_3^B)}
\end{equation}
with $\zeta_N=e^{2\pi i/N}$.
Repeating the group average produces
\begin{equation}
\begin{split}
(X_p^A)^k &\rightarrow (X_p^A)^k\squad,\\
(X_p^B)^k &\rightarrow (X_p^B)^k\,\delta^{(N)}_{k\Delta n_p^C}\squad,\\
(X_p^C)^k &\rightarrow (X_p^C)^k\,\delta^{(N)}_{k\Delta n_p^B}
\end{split}
\end{equation}
for $k\in\{1,\ldots,N-1\}$. As a result, the nontrivial single-chain sector has the form
\cite{spt_zN3}
\begin{equation}
\widetilde H_{\partial}^{(N)}=-\sum_p\sum_{k=1}^{N-1}
X_p^k\,\delta^{(N)}_{k(n_{p+1}-n_{p-1})} \squad.
\end{equation}
For $N=4$ this constrained form  indeed
has the same structure as in Eq.~\eqref{eq:ham_init},
but the arithmetic of $N$ controls the projectors \cite{spt_zN3}:
if $N$ is prime, every nonzero $k$ is invertible and
$\delta^{(N)}_{k\Delta n}=\delta^{(N)}_{\Delta n}$; if $N$ is composite, a term with
$d=\gcd(k,N)>1$ imposes only $\Delta n \equiv 0\;({\rm mod}\;N/d)$ and not $\Delta n=0$
(here, $\gcd(k,N)$ stands for ``greatest common divisor": the largest positive integer that divides both $k$ and $N$ without a remainder).

\section{Numerical analysis of the boundary model}

The edge Hamiltonian \eqref{V2H} can be presented in a simple form
\begin{equation}
\begin{split}
\hspace{-1pt}\widetilde{H}_\partial = -\frac{1}{4}\sum_p\Big[&X_p +
Z_{p-1} (X_p + X_p^\dagger) Z_{p+1}^\dagger \\
+ &X_p^2+Z_{p-1}^2(X_p+X_p^2) Z^{\dagger 2}_{p+1} + \text{h.c.}\Big] \squad.
\end{split}
\label{em-1}
\end{equation}

\begin{figure}
	\centering
	\includegraphics[width = .9\linewidth ]{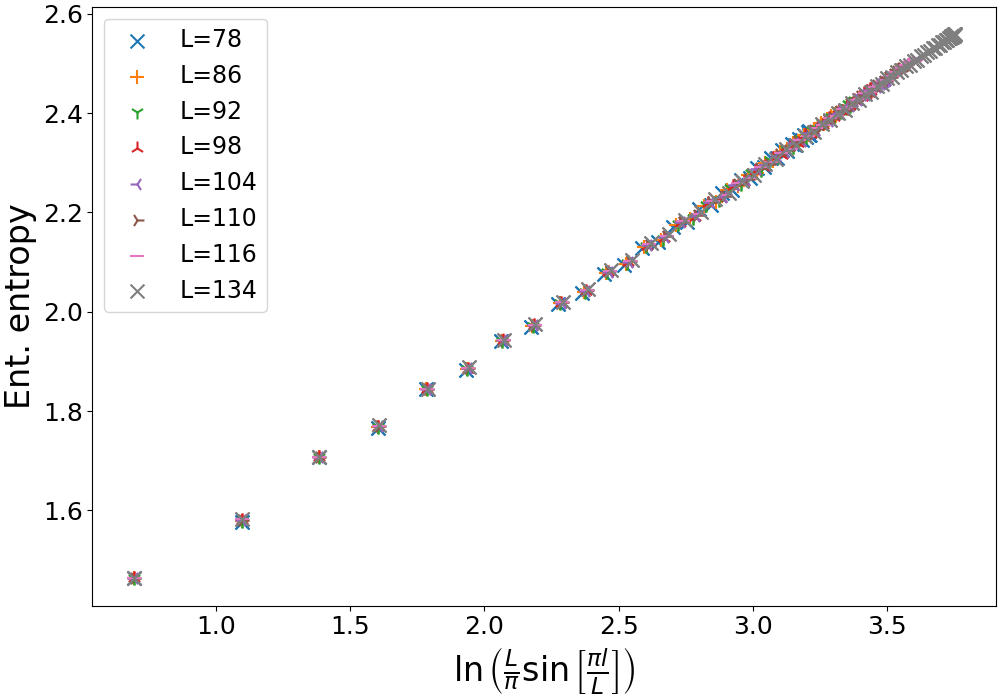}
	\caption{The entanglement entropy on reparameterized subsystem size ratio $l/L$ for various system sizes $L$. Central charge $c=2.191 \pm 0.004$ can be determined from the slope.}
	\label{fig:entropy}
\end{figure}

According to \cite{Calabrese-2009}, the entanglement entropy
in a conformal field theory with central charge $c$ on an open system of length
$L$ is given by
\begin{equation}
\label{eq:EE-1}
S_L(l)= a + \frac{c}{6}\ln\inb({\frac{L}{\pi} \sin\inb[{\frac{\pi l}{L}}]})\squad,
\end{equation}
where $l$ is the length of the considered subsystem and $a$ is some constant.
For a closed system, the coefficient $c/6$ in front of the $\log$ should be replaced by $c/3$. 
This formula defines the finite-size behavior of the entanglement entropy in a chain model at criticality.
 Therefore, it allows us to extract the central charge $c$ from the finite-size scaling of the entanglement entropy.

Running DMRG (which is used in this study) on periodic chains for this edge model is
much more costly in terms of computational resources, both in runtime and memory,
compared to the equivalent calculations on open chains.
Here we have concentrated on open chains to reach larger sizes and the desired precision.

The data in \reff{entropy} shows the numerically obtained values of
entanglement entropy versus the rescaled subsystem size $l$, 
$\ln\inb({\frac{L}{\pi} \sin\inb[{\frac{\pi l}{L}}]})$ for various 
$L=78,86,92,98,104,110,116,134$ for an open chain \cite{data}.
We see a perfectly straight line for all $L$-s in agreement with \eqref{eq:EE-1},
from which the values of parameters $c=2.191 \pm 0.004 \simeq 11/5$ and $a= 1.186 $
are obtained.

We then analyze the spectrum of low-energy modes of the derived boundary Hamiltonian
to check for gapless excitations,
which are a prerequisite for describing the system within a conformal field theory paradigm.
The DMRG simulations of the edge Hamiltonian provide strong evidence that the boundary theory is indeed gapless.

\begin{figure}
	\centering
	\includegraphics[width = .9\linewidth ]{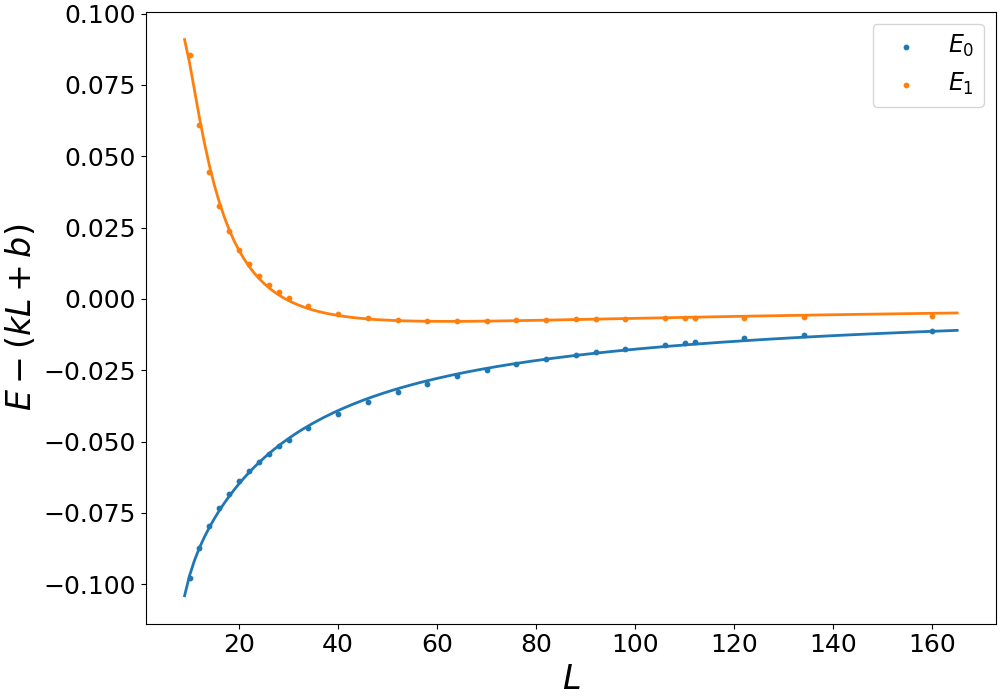}
	\caption{The ground state energy $E_0$ and the first excited state energy $E_1$ on system size $L$, after subtracting the linear term in $L$. The numerical results (dots) and the corresponding fitting curves. $h^{(\mathrm{b})}_1=(0.019 \pm 0.009) c$.}
	\label{fig:gap}
\end{figure}

\reff{gap} displays the ground state energy $E_0$
and the first excited state energy $E_1$ versus the system size $L$ for open chains.
According to ~\cite{Cardy-1986-a,Cardy-1986,cardy-scaling,oshikawa-scaling}
the finite-size  behavior of the
lower energy states read as
\begin{equation}
\begin{split}
    &E_0(L) = k L + b - \frac{\pi v c}{24L} + {\cal O}(L^{-2}) \squad,\\
    &E_i(L) = E_0(L)+\frac{\pi v}{L}h^{(\mathrm{b})}_i+{\cal O}(L^{-2}) \squad,
\label{energies}
\end{split}
\end{equation}
where $k, b, v$ are some constants, $c$ is the central charge and $h^{(\mathrm{b})}_i$
are the \emph{boundary} scaling dimensions (a single chiral sector) set by the conformal boundary conditions of the microscopic edge.
The decomposition into bulk weights $(h_i,\bar h_i)$ and the momentum quantization $P\propto (h_i-\bar h_i)$
apply to periodic chains.

The excitation gap is defined as the energy difference between the first excited state and the ground state, $\Delta^{(\mathrm{b})}_1=E_1-E_0$.  As a function of system size $L$, this finite-size gap decreases steadily as $L$ grows,
\begin{equation}
\label{gap}
\Delta^{(\mathrm{b})}_1 \simeq \frac{\pi v h^{(\mathrm{b})}_1}{L} \squad.
\end{equation}
Fitting the numerical data for $E_0(L)$ and  $E_1(L)$ by the functions \eqref{energies}
and using the values for the central charge $c= 2.191 \pm 0.004$, we have  obtained the numerical value
for the smallest boundary scaling dimension accessible in the open-chain spectrum via DMRG within the symmetry sector
to be $h^{(\mathrm{b})}_1 = 0.04 \pm 0.02$, within the available precision.
Directly extracting the bulk weights $(h,\bar h)$ and the momentum quantization $P\propto (h_i-\bar h_i)$ requires periodic boundary conditions and momentum resolution.
The presented calculations have been done for an open system of sizes $L= 10 \sim 160 $.



Motivated by the $\mathbb{Z}_3$ and $\mathbb{Z}_3^{\times 3}$ SPT edges \cite{spt_z33,spt_z3}, it is natural to ask whether the present $\mathbb{Z}_4^{\times 3}$ edge criticality is also described by coset CFT $SU_k(3)/SU_k(2)$ at some level $k$. A particularly suggestive possibility is at $k=3$, since the numerically extracted central charge $c=2.191 \pm 0.004$ is close to the rational value $11/5$. 

For the Wess-Zumino-Novikov-Witten models one has
\begin{align}
&c\left(SU(N)_k\right)=\frac{k (N^2-1)}{k+N} \squad,\label{coset-c}\\
&c_{\text{coset}}(k)=c\inb({SU(3)_k})-c\inb({SU(2)_k})=\frac{8k}{k+3}-\frac{3k}{k+2}\squad.
\nonumber
\end{align}
Setting $k=3$ gives $c_{\text{coset}}(3)=11/5$, which agrees with our entanglement entropy estimate within the numerical precision. This is strong evidence that low-energy edge theory is compatible with the $SU(3)_3/SU(2)_3$ coset.

The definitive identification, however, requires matching the operator content based on the conformal towers\cite{DiFrancesco1997}, not only $c$. 
Our preliminary open-chain estimate for the \emph{boundary} scaling dimension in a single chiral sector, determined by the conformal boundary conditions of the microscopic edge, is $h^{(\mathrm{b})}_1 = 0.04 \pm 0.02$.
It does not obviously coincide with one of the simplest values quoted above. This may reflect a finite-size effect, a symmetry-sector restriction in DMRG computation, or the specific conformal boundary condition realized by $\widetilde{H}_\partial$. Therefore, to sharpen the CFT identification beyond the central charge, one will need to:
(i) extract several low-lying gaps (and degeneracies) in symmetry-resolved sectors and match them to conformal towers,
(ii) perform smaller-size periodic-boundary calculations to access both energies and momenta, enabling a direct extraction of bulk $(h,\bar h)$, and (iii) analyze correlators and/or the entanglement spectrum to identify the leading lattice operators and their scaling dimensions.
Our preliminary data of several low-lying excited states on shorter open chains did not show anomalous convergence problems, including in near-degenerate cases. However,
the available sizes are not yet sufficient for a stable assignment of several scaling dimensions.
This indicates that the conformal-tower analysis is technically feasible, but that reliable extraction
of bulk weights $(h_i,\bar h_i)$ and momentum quantization $P\propto(h_i-\bar h_i)$ will require
periodic-boundary calculations with momentum resolution, which are substantially more demanding than
the open-chain DMRG calculations reported here.

\section{Conclusion}
We have studied $\mathbb{Z}_4^{\times 3}$ Potts model in paramagnetic phase and constructed one set of its seven $\mathbb{Z}_4^{\times 7} $ SPT phases defined by the cohomology group $H^3(\mathbb{Z}_4^{\times 3}, U(1))=\mathbb{Z}_4^{\times 7}$.
In particular, we focused on the diagonal action of the symmetry, where each site of the triangular lattice hosts three Potts variables—each transforming under a different copy of $\mathbb{Z}_4$. Using this setup, we explicitly constructed the corresponding $\mathbb{Z}_4$
 cocycles that generate all SPT phases of $\mathbb{Z}_4$. More precisely, we focused on the antisymmetrized  $\omega_3^a$ in Eq.~(\ref{wa}),
which is designed to satisfy the antisymmetry constraint Eq.~(\ref{nu-asym}) needed for a shape-independent edge Hamiltonian. As indicated by $R[\omega_3^a]=-1$, this representative is naturally interpreted as an order-two element within the
``colorless" $\mathbb{Z}_4$ factor, and
extending the construction to a full set of seven independent generators of
$H^3(\mathbb{Z}_4^{\times 3},U(1))$ (and to other order-four representatives)
is left for future work.

 We have also formulated a gapless edge-state chain model with symmetry $\mathbb{Z}_4$, which is a hallmark of a nontrivial SPT phase. Numerically, we found that the central charge of these edge states is $c\simeq 11/5$ and the boundary scaling dimension in a single chiral sector is $h^{(\mathrm{b})}_1 \approx 0.04$. Our central charge estimate is compatible with the $SU(3)_3/SU(2)_3$ coset CFT, which has an exact $c=11/5$. Verifying the continuum theory beyond the central charge involves a systematic comparison of the low-energy spectrum with the expected conformal towers, including symmetry-resolved sectors and degeneracies, and ideally additional periodic-boundary data. Finally, since our gap analysis is based on an open chain, the exponent we find should be considered a boundary scaling dimension; identifying $(h,\bar h)$ in the bulk requires periodic spectra with momentum resolution.

 \section*{Acknowledgments}
  The authors acknowledge  Armenian HESC grants 24RL-1C024  (HT, TS), 21AG-1C024 (MM,AS),  24FP-1F039 (TH, HT, AS), and 21AG-1C047(TH) for financial support.

\bibliography{refs-z4}

\end{document}